\documentclass[twocolumn,showpacs,preprintnumbers,amsmath,amssymb]{revtex4}

\usepackage{graphicx}
\usepackage{dcolumn}
\usepackage{bm}

\begin{document}


\title{An effective long-range attraction between protein molecules in solutions studied by small angle neutron scattering}

\author{Yun Liu$^1$, Emiliano Fratini$^2$,
Piero Baglioni$^{1,2}$, Wei-Ren Chen$^1$, Sow-Hsin
Chen$^1$\footnote{To whom correspondence should be addressed.
Email: sowhsin@mit.edu} }
\affiliation{$^1$Department of Nuclear Engineering, Massachusetts Institute of Technology, Cambridge, MA 02139. \\
$^2$Department of Chemistry and CSGI, University of Florence, via
della Lastruccia 3, 50019 Florence, Italy}
\date{\today}

\begin{abstract}
Small angle neutron scattering intensity distributions taken from
cytochrome C and lysozyme protein solutions show a
rising intensity at very small wave vector, Q, which can be
interpreted in terms of the presence of a weak long-range
attraction between protein molecules. This interaction has a range
several times that of the diameter of the protein molecule, much
greater than the range of the screened electrostatic repulsion. We
show evidence that this long-range attraction is closely related
to the type of anion present and ion concentration in the
solution.
\end{abstract}

\maketitle

The bottleneck of protein crystallography is the lack of
systematic methods to obtain protein crystals. This is partly due
to incomplete understanding of the physical chemistry conditions
controlling the growth of protein crystals. A full
comprehension of the effective protein interactions and phase
behavior is therefore essential. It has been shown that the
crystallization curves of some globular proteins appear to
coincide with the phase diagrams of a hard sphere system
interacting with a short range attraction \cite{Rosenbaum, Hagen,
Pellicane}. Small angle neutron and X-ray
scattering investigations of proteins suggest the presence of a short-range attractive
interaction between protein molecules besides the electrostatic
repulsion induced by the residual charges \cite{Tardieu1999,
Stradner, Baglioni}. The
DLVO potential has been successfully applied to many
colloidal systems and  protein
solutions \cite{Tardieu1999, Pellicane}. However, it
does not seem to fully explain the rich behaviour of proteins
\cite{Tardieu1999,Broide,Piazza,Striolo}, and due to the complexity
of these systems (anisotropic property, irregular shape, distributed charge
patches, etc.), a complete understanding of the properties of the
effective interactions between protein molecules in solutions
remains a challenge \cite{Piazza}.

Recent measurements of small angle neutron scattering (SANS)
intensity distribution in protein solutions show interesting
results \cite{Stradner, Baglioni,Baglioni2}. Beside the
normal first diffraction peak, it is present a peak (cluster peak)
appearing at a much smaller scattering wave vector, $Q$, due to
the formation of ordered clusters. The appearance
of a cluster peak is explained as due to the competition of a
short-range attraction and a long-range electrostatic repulsion
\cite{Stradner, Sciortino, Liu_JCP}. Moreover, a rising intensity
as $Q$ approaches zero (zero-$Q$ peak) is observed in both
liquid-like and solid-like samples, which implies that the
effective potential should have more features in addition to the
well known short-range attraction and electrostatic repulsion.
The existence of a long-range attraction between protein molecules
has already been postulated in theoretical studies.
Noro et al. suggested that the presence of a long-range attractive
force between protein molecules should shift the metastable
fluid-fluid critical point out of gel regime \cite{Noro}. Thus a
protein crystallization may occur without gelation.
By employing two Yukawa potential model, Lawlor et al. showed that
the introduction of a long-range attraction between protein
molecules can enhance crystal growth by avoiding the formation of
a disordered state, an attractive glass \cite{dawson2,chen}, while
preserving the equilibrium features of a system with the
short-range attraction \cite{dawson1}.

In this paper, by systematically studying the zero-$Q$ peak, we
find that a weak long-range attraction needs to be considered to
explain the SANS scattering intensity distributions. The properties of this
long-range attraction potential are also investigated.
Cytochrome C from horse heart (product no. C7752) was purchased
from Sigma and is obtained using a procedure that avoids the
trichloroacetic acid (TCA) that is known to promote the dimer
formation. Cytochrome C has been
dialyzed three times in order to remove any extra salt. Buffers
are not used to avoid possible bindings of organic molecules to the protein surface.
Lysozyme from chicken egg white was purchased from Fluka (product
no. L7651) and used without further purification. The pD of cytochrome solutions
at 7.2 and 9.5, has been adjusted by a HCl standard solution. Only samples of
Lysozyme have been prepared in 20 mM HEPES buffer. The pD value
was sequentially checked by an ISFET pHMeter (KS723) before and
after each performed experiment and found stable within ±0.1
units. All samples were prepared a few days before the scheduled
experiments to allow the hydrogen/deuterium exchange. For all solutions the final
protein concentration has been measured by UV-Visible
spectroscopy.
Our experiments were performed at the small angle neutron
scattering station, NG7, at the Center of Neutron Research in the
National Institute of Standard and Technology. Two configurations
have been used to reach a wide range of the wave vector, $Q$, from
$0.004 \AA^{-1}$ to $0.30 \AA^{-1}$, where $Q=\frac{4\pi}{
\lambda} sin(\theta/2)$, $\lambda$ is the neutron
wavelength, $\theta$ the neutron scattering angle. 
All the analyses have taken into account the instrumental
resolution correction.

SANS intensity distribution, $I(Q)$, can be expressed as
$I(Q)=A\overline{ P}(Q) S(Q)$, where $\overline{P}(Q)$ is the
normalized particle structure factor, $S(Q)$, the inter-particle
structure factor, $A$, the known amplitude factor which is
proportional to the volume fraction and the square of the neutron
scattering length contrast between protein and solvent \cite{Wu}.
Cytochrome C has an ellipsoidal shape with the semi-major and
minor axes, $a\times b\times b=15 \times 17 \times 17 \AA^3$,
while a lysozyme molecule has a dimension $a\times b\times b=22.5
\times 15 \times 15 \AA^3$.
$\overline{P}(Q)$ is calculated by considering the
ellipsoidal shape of the protein. 
$S(Q)$ is calculated by solving the Ornstein-Zernike (OZ) equation
within the mean spherical approximation (MSA) closure involving an
effective pair potential, $V(r)$, to be specified later. The
effect on $S(Q)$ due to the non-spherical shape of a particle is approximately
taken into account by using the decoupling
approximation \cite{Kotlarchyk}.

 \begin{figure}
   \includegraphics[width=7.5cm]{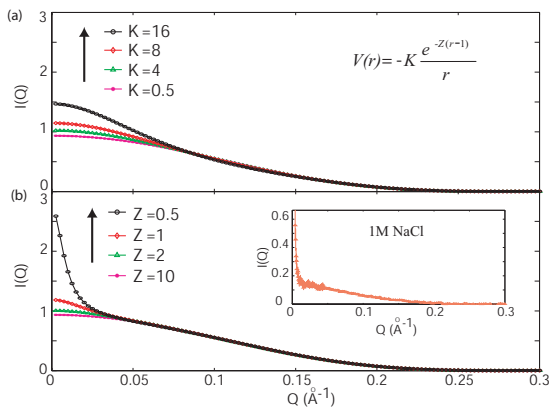}
   \caption{Theoretical calculations of $I(Q)$ resulting from one Yukawa attraction at $1\%$
    volume fraction.
   a)  the effect due to the variation of the attraction strength, $K$ at $Z=10$.
   b) the effect due to the variation of the attraction range,
   $1/Z$ at $K=0.5$. The inset shows $I(Q)$ from a cytochrome C sample at $1\%$ volume fraction
   in $1M~NaCl$ at pD=11.}
   \label{}
 \end{figure}

Figure 1 shows theoretical calculations together with the SANS
result from a cytchrome C solution at $pD=11$ with $1\%$ volume
fraction in $1M$ $NaCl$ in the inset. Conventionally, the
effective inter-protein potential is considered to consist of a
short-range attractive and a long-range electrostatic repulsive
part. The repulsion is screened out by the concentrated salt.
Therefore, $S(Q)$ could be obtained by solving OZ
equation by considering only a short-range attraction, which can
be approximated by an attractive potential of a Yukawa form
\cite{Malfois}, $V(r)=-K \frac{e^{-Z(r-1)}}{r}$, where $K$ is
normalized by $k_BT$, and $r$ is the inter-protein distance
normalized by $\sigma$ with $\sigma=2(ab^2)^{1/3}$. The
interaction range, $\frac{1}{Z}$, is approximately about $10\%$ of
the protein diameter\cite{Tardieu1999}. Figure 1(a) shows the
theoretically calculated SANS intensity distribution with $Z=10$
by taking the amplitude factor $A$ equal to unity.
As $K$ increases from $0.5$ to $16$, the intensity at $Q=0$ increases gradually.
The results from a short-range attraction always smoothly change
the whole scattering curve and can not reproduce the sharp rising
up of experimental SANS intensity at very low $Q$ (zero-$Q$ peak)
as shown in the inset. Since the typical strength of short-range
attraction is smaller than $10 k_BT$\cite{Tardieu1999, Eberstein,
Xia}, the short-range attraction can not explain the observed SANS
intensity distribution. In Figure 1(b), $K$ is fixed at $0.5$.
When $Z$ decreases from $10$ to $0.5$,
i.e. the attraction range increases, the intensity at low $Q$
increases sharply and the theoretical curve exhibits a similar trend
to the experimental intensity. The comparison between the
experimental result and theoretical curves thus suggests that the
zero-Q peak could be induced by a weak long-range attraction.
Therefore, the effective potential between protein molecules in
solutions should consists of three features: a short-range
attraction, an intermediate-range electrostatic repulsion, and a
weak long-range attraction. This zero-$Q$ peak has been overlooked
in previous experiments due to the limited $Q$ range covered
\cite{Wu}. However, it has been observed before in lysozyme protein
solutions \cite{Giordano} and was explained as due to the
long-range density fluctuation.

 \begin{figure}
   \includegraphics[width=7cm]{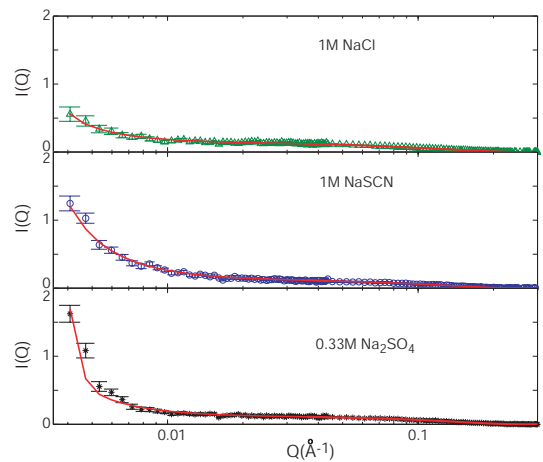}
   \caption{SANS intensity distribution of cytochrome C solutions at pD=11
   and at $1\%$ volume fraction with different salts added. The solid lines are
   the theoretical analyses.}
   \label{}
 \end{figure}

To investigate the properties of this long-range attraction, three
different cytochrome C samples at pD=11 with $1\%$ volume fraction
added with different salts were measured and their SANS intensity
distributions with error bars are plotted in semi-log scale to
clearly show zero-$Q$ peaks in Figure 2.
The salt concentration, which is indicated in the figure, has been
changed to keep the same ionic strength. Very interestingly, the
results show that the zero-$Q$ peak depends on the different
anions added. $NaCl$ induces a much weaker zero-$Q$ peak compared
with that of $NaSCN$ and $Na_2SO_4$. Since the true form of this
long-range attraction is still unknown, in order to fit the
experimental results, we assume that the effective inter-protein
potential can be simulated by two attractive Yukawa form
potential,
$V_{TY}(r)=-K_1\frac{e^{-Z_1(r-1)}}{r}-K_2\frac{e^{-Z_2(r-1)}}{r}$.
The first term is used to simulate the short-range attraction,
while the other one is used to simulate the long-range attraction.
The fitted results are given in table I.

\begin{table}
\caption{Fitted parameters using the two attractive Yukawa form
potential. Results are shown in Figure 2.}
\begin{ruledtabular}
\begin{tabular}{cccccc}
       & $a(\AA)$ & $K_1$ & $Z_1$ & $K_2$ & $Z_2$\\
 $NaCl$&14.7& $7\pm 3$ & $7\pm 3$ & $0.11\pm 0.02$ & $0.20\pm 0.01$ \\
 $NaSCN$    &  14.6    & $8\pm 5$ & $10\pm 6$ & $0.33\pm 0.02$ &
 $0.35 \pm 0.01 $ \\
 $Na_2SO_4$ &  14.7    & $4\pm 5$ & $10 \pm 10$ & $0.24 \pm 0.02$ &
 $0.27 \pm 0.02$ \\
\end{tabular}
\end{ruledtabular}
\end{table}

The fitting is not sensitive at all to the short-range attraction.
The long-range attraction has a weak strength (less than
$0.5~k_BT$) and a range of $3 \sim 5$ times the protein diameter,
and it is very sensitive to the anions added to the solutions. This
implies that the long-range attraction is at least partly induced
by the ion cloud around protein molecules. Depletion forces and
van der Waals force can not explain this feature. It is very interesting
to notice that the strength and range of the long-range attraction is
about the same value as the attractive potential between
like-charged particles \cite{Sogami}.
However, a charged colloidal particle typically has a uniform
charge distribution on the surface, while a protein molecule has
both positive and negative charge patches, which make the ion
cloud distribution around a protein much more complicated.

 \begin{figure}
   \includegraphics[width=7cm]{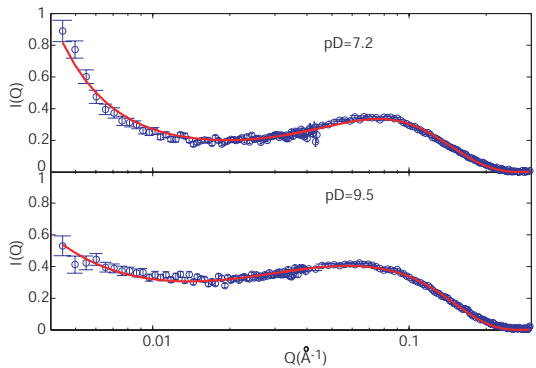}
   \caption{SANS intensity distributions of cytochrome C samples at $5\%$
   volume fraction. The solid lines are theoretical analyses using
   two Yukawa potential.}
   \label{}
 \end{figure}

Since the long-range attraction seems to be induced by the ion
cloud, if we can minimize the ion concentration in solutions,
we should expect to greatly suppress the zero-$Q$ peak. Figure 3
shows SANS results of two cytochrome C samples at pD=7.2 and
pD=9.5 with $5\%$ volume fraction. Without adding any extra salt in solutions,
the ion concentration is determined by the dissociated charges
from proteins.
The cytochrome C molecule has much smaller charge number at pD=9.5 than
that at pD=7.2, being the isoelectric point, $pI$, about 10.2. This leads to
a smaller ion concentration in solutions at pD=9.5. Correspondingly, the
SANS intensity distribution shows a zero-$Q$ peak weaker than that
at pD=7.2. This implies that the zero-Q peak depends on the ion concentrations,
and  is not likely due to permanent
cluster formations, since a weaker repulsion should favor larger
cluster formations and thus induce larger zero-$Q$ peak at pD=9.5,
which is in contrast with our experimental results.
In order to quantitatively analyze the zero-$Q$ peak, we again
used the two-Yukawa potential model. However, due to the existence
of the weak long-range attraction, we need three Yukawa terms to
completely simulate the features of the potential. We argue that
in cytochrome C solutions, the short-range attraction is small so
that its effect on the structure factor should be very small at
the relatively low volume fraction. Actually, the first
diffraction peak of SANS intensity distribution in cytochrome C
solutions has been successfully analyzed by only considering the
electrostatic repulsion at various pD values \cite{Wu, Liu_JCP}.
Furthermore, for cytochrome C solutions, we did not observe the
cluster peak as it is observed for lysozyme solutions.
The lack of cluster peak is attributed to the weak short-range
attraction. Therefore, we ignore the short-range attraction and we
use the second term of $V_{TY}$ for the long-range attraction
contribution and the first term of $V_{TY}$ to simulate the
electrostatic repulsion, which can be calculated by the charge
number of the protein molecule and the ionic strength \cite{Wu,
Liu_JCP}. The fitted results given in table II show that the range
of the long-range attraction is about $3 \sim 5$ times of protein
diameter, and also confirm our direct observation from Figure 3,
i.e. the strength of long-range attraction at pD=9.5 is much
smaller.

\begin{table}
\caption{Fitted parameters obtained using the two Yukawa potential. Results
are shown in Figure 3.}
\begin{ruledtabular}
\begin{tabular}{ccccc}
       & $a(\AA)$ & charge number & $K_2$ & $Z_2$\\
 $pD=7.2$&15.3& $3.8\pm 0.1$ & $0.37\pm 0.02$ & $0.34\pm 0.01$ \\
 $pD=9.5$&15.4& $1.7\pm 0.1$ & $0.08\pm 0.01$ & $0.21 \pm 0.01 $ \\
\end{tabular}
\end{ruledtabular}
\end{table}

 \begin{figure}
   \includegraphics[width=7.5cm]{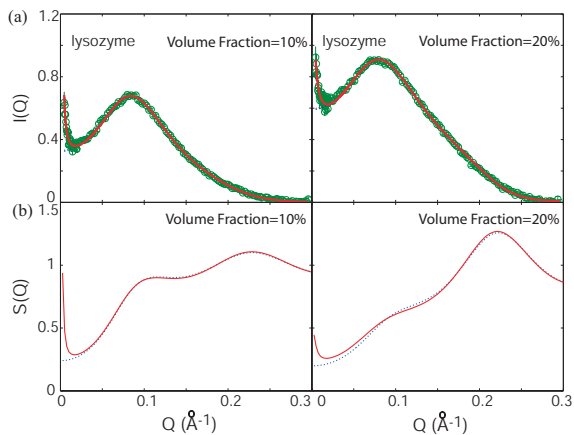}
   \caption{SANS intensity distributions of lysozyme samples at pD=5.1 with
   $20~mM$ HEPES buffer. Dotted lines (a and b) are fitted by considering
   only a short-range attraction and electrostatic repulsion. Solid lines (a and b)
   are the fitted results by using three Yukawa form potential. Lower panel
   shows the calculated structure factor, $S(Q)$.}
   \label{}
 \end{figure}

After showing the existence of the long-range attraction in
solutions dominated by monomers, it is important to check its
existence in protein solutions having equilibrium clusters. Figure
4 shows the results of lysozyme protein solutions at pD=5.1 and at
$10\%$ and $20\%$ volume fraction in HEPES buffer. The top panel
shows SANS intensity distributions with the fitted results and the
bottom panel shows the calculated $S(Q)$ from the fitted
parameters. The {\it main peak} in the top panel is a cluster peak
as it is independent of volume fraction.
Interestingly, even in the presence of cluster peak, both results
have shown the zero-$Q$ peak, which was overlooked by previous
experiments \cite{Stradner}. The coexistence of the cluster peak
and zero-$Q$ peak clearly indicates the necessity of introducing a
third potential feature, a long-range attraction. We used the two
Yukawa model previously described to fit the main peak, assuming
that the first term is the short range attraction, and the second term
the electrostatic repulsion. The fitted curves are shown as dotted
line in the figure and the fitting results are given in table III.
\begin{table}
\caption{Fitted parameters from fitting the cluster peaks by
using only a short-range attraction and the electrostatic repulsion (see
Figure 4).}
\begin{ruledtabular}
\begin{tabular}{ccccc}
       & $a(\AA)$ & charge number & $K_1$ & $Z_1$\\
 $10\%$&21.1& $7.0\pm 1.2 $ & $6.9\pm 1.0$ & $14\pm 5$ \\
 $20\%$&21.2& $6.8\pm 1.5$ & $7.7\pm 1.2$ & $15 \pm 5 $ \\
\end{tabular}
\end{ruledtabular}
\end{table}
The attraction strength is about $7k_BT$ with $7\%$ of attraction
range, which is consistent with Ref.\cite{Eberstein}. In order to
fit the complete SANS distribution, we use the two Yukawa model as
the reference system and treat the long-range attraction as a
perturbation by employing the random phase approximation
\cite{Shukla}. The long-range attraction is considered here as a
third Yukawa form, $-K_3\frac{e^{-Z_3(r-1)}}{r}$. The results
fitted with this approximation are shown as solid line. The
addition of this third potential feature slightly changes the
previous fitting parameters. The parameters for the long-range
attraction are, $K_3=0.038\pm 0.005$ and $Z_3=0.17\pm 0.01$ for
the $10\%$ lysozyme sample, while for $20\%$ sample, $K_3=0.019\pm
0.002$ and $Z_3=0.21\pm 0.03$.

Judging from SANS results on both cytochrome C and lysozyme
protein solutions, we believe that the existence of weak
long-range attraction is universal for all protein solutions. Its
strength and range depend both on the type of anion and ion
concentration in solutions. In other words, it depends on the ion
cloud around protein molecules. Our finding has important
implications in understanding the protein crystallization process.
There are numerous measurements of second virial coefficient in
protein solutions by static light scattering experiments
\cite{George, Velev}. Due to the very small $Q$ range the light
scattering can access, its intensity is closely related to the
height of the zero-$Q$ peak, which depends on the feature of the
long-range attraction. Thus we believe that the conclusions
derived from those observations should take into account the
existence of the weak long-range attraction. The charge dependence
of some of these phenomena can then be naturally related to the
charge dependence of the long-range attraction \cite{Muschol,
Piazza}. Also the very large cluster formation or gelation in
supersaturated protein solutions may be induced by the long-range
attraction \cite{Sokuri, Niimura}.

We acknowledge a grant support from Materials Science Division of
US DOE, DE-FG$02$-90ER$45429$. We are indebted to the NIST Center
for Neutron Research for providing neutron scattering facilities
used in this work. We profitted from being affiliated with the EU
funded Marie-Curie Research Network on Arrested Matter. EF and PB
acknowledge MIUR (PRIN-2003 grant) and
CSGI(Florence, Italy) for partial financial support.


\end{document}